# Near octave-spanning perfect soliton crystals in AlN microresonators

HAIZHONG WENG,[1] ADNAN ALI AFRIDI,[1] JIA LIU,[2] JING LI,[1] JIANGNAN DAI,[2] XIANG MA,[2] YI ZHANG,[2] QIAOYIN LU,[2] WEIHUA GUO[2,3], AND JOHN F. DONEGAN[1,4]

[1]*School of Physics, CRANN and AMBER, Trinity College Dublin, Dublin 2, Ireland*
[2]*Wuhan National Laboratory for Optoelectronics, and School of Optical and Electronic Information, Huazhong University of Science and Technology, 1037 Luoyu Road, Wuhan 430074, China*
[3]*e-mail: guow@mail.hust.edu.cn*
[4]*e-mail: jdonegan@tcd.ie*

**Dated: February 20, 2021**

**The perfect soliton crystal (PSC) was recently discovered as an extraordinary Kerr soliton state with regularly distributed soliton pulses and enhanced comb line power spaced by multiples of the cavity free spectral ranges (FSRs). The modulation of continuous-wave excitation in optical microresonators and the tunable repetition rate characteristic will significantly enhance and extend the application potential of soliton microcombs for self-referencing comb source, terahertz wave generation, and arbitrary waveform generation. However, the reported PSC spectrum is generally narrow. Here, we demonstrate the deterministic accessing of versatile perfect soliton crystals in the AlN microresonators (FSR ~374 GHz), featuring a broad spectral range up to 0.96 of an octave-span (1170-2300 nm) and terahertz repetition rates (up to ~1.87 THz). The measured 60-fs short pulses and low-noise characteristics confirms the high coherence of the PSCs.**

Since dissipative Kerr solitons (DKSs) were first observed in a MgF$_2$ microresonator [1] through careful balancing of Kerr nonlinearity and dispersion, a great number of researchers have devoted time to produce soliton microcombs in various platforms such as silica [2], Si$_3$N$_4$ [3-7], AlN [8], LiNbO$_3$ [9, 10], AlGaAs [11]. Due to the merits of miniaturization, high coherence, broadband spectral range, and a repetition rate in the microwave regime, Kerr soliton combs are promising for a wide range of applications such as optical clock, frequency synthesizer, coherent communications, astronomical spectrometer calibration, dual-comb spectroscopy, and microwave photonics [12-14]. Subsequently, novel soliton types such as breather solitons [15, 16], Stokes solitons [17], and soliton molecules [18] were also demonstrated in various microresonators.

The soliton crystal (SC) is a specific soliton state with clearly discrete comb lines spaced by multiple cavity free spectral ranges (FSRs) [19, 20]. Such SCs are typically formed in the presence of avoided mode crossing (AMX), which induces a modulation on the intracavity continuous-wave (CW) background, leading to the interference between the solitons [20]. The AMX was initially utilized to alter the localized dispersion thus realizing a dark pulse in normal-dispersion microring resonators (MRRs) [21, 22]. Among the variety of SC states, the perfect soliton crystal ($m$-PSC) is one ideal representative, which has $m$ soliton pulses uniformly distributed in the cavity. In contrast to a single soliton, the PSCs feature high conversion efficiency and $m^2$ enhanced comb power [23, 24]. Therefore, the PSCs can break away from the repetition rate limitation and extend microcomb applications to the THz regime while avoiding resonators with a small radius. As revealed in [24, 25], a relatively low pump power is helpful to access the PSCs, which may in turn restrict the spectral bandwidth. As far as we know, the broadest band PSC ($m$=3, 1380-1870 nm) in the near infrared regime was formed in a Si$_3$N$_4$ MRR with an FSR of ~1 THz [24].

Due to the strong Kerr ($\chi^3$) and Pockels ($\chi^2$) nonlinearities, high $Q$ AlN MRRs were employed to produce broadband microcombs and Kerr solitons [8, 26-29], as well as the near-visible microcombs [30] and Pockels solitons [31]. Most recently, we have realized the direct accessing of an octave-spanning single-DKS in an optimally designed AlN MRR [32], which will eliminate the complicated control or extra equipment often required for soliton formation.

Here, we demonstrate the deterministic generation of versatile PSCs in AlN MRRs due to the diverse mode interactions. The high intrinsic $Q$ factor ($Q_{int}$), low anomalous dispersion and pulley waveguide coupling together promote the formation of a PSC with a near octave-spanning spectral range ($m$=4, 1170-2300 nm). The switching between different PSCs states is also obtained by simply tuning the pump wavelength.

As the schematic presented in Fig. 1(a) shows, the PSCs with various soliton numbers are excited in different AlN MRRs, coupled with straight or pulley waveguide. The MRRs are patterned from an AlN-on-sapphire wafer (AlN thickness 1.2 μm) via standard photolithography and inductively coupled plasma etching processes [33]. All the resonators employed in this work have a radius of 60 μm and a cross section of 2.29×1.2 μm² (fully etched), which were designed to ensure a low anomalous dispersion ($D_2/2\pi$ = 4.8 MHz) of the target fundamental transverse electrical (TE$_{00}$) mode [32]. The bus waveguide width $W$ and coupling gap $G$ were designed to vary from device to device (see details in Supplement 1) in an attempt to produce resonators with a controllable $m$×FSR. Figure 1(b) shows the transmission of resonator 1 ($W$= 0.91 μm, $G$ = 0.5 μm) with a fiber-to-chip coupling loss of ~3.2 dB per facet. The TE$_{00}$ and first-order TE (TE$_{10}$) modes are close to each other near 1550.7 nm ($\mu$=0), accompanied by the significantly changed linewidths and a dramatic decrease of the extinction ratios at $\mu$ = -1, illustrating a weak coupling between the two transverse modes. The normalized TE$_{00}$ resonance at 1565.73 nm ($\mu$ = -5), used for PSC generation, is depicted in Fig. 1(d) with a loaded $Q$ factor ($Q_{load}$) and $Q_{int}$ of ~5.2×10$^5$ and ~1.3×10$^6$, corresponding to a propagation loss of 0.28 dB/cm.

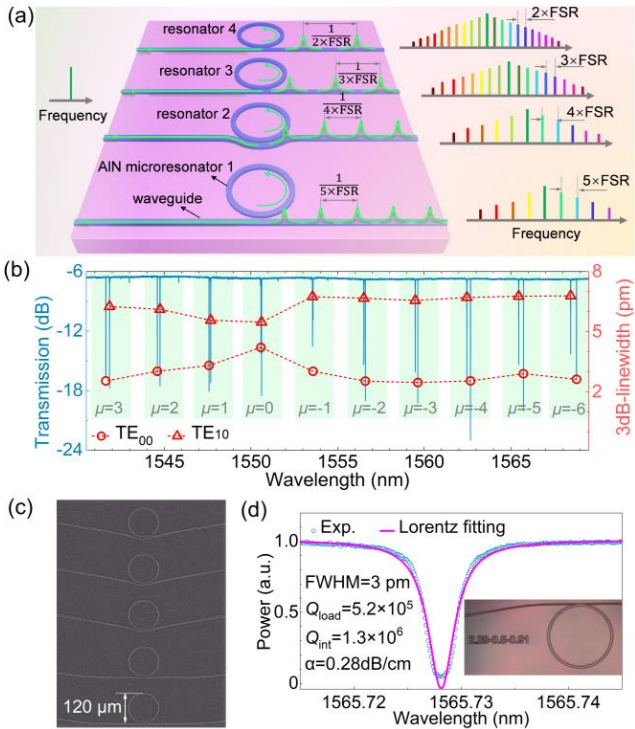

**Fig. 1.** (a) Schematic of PSCs generation based in an AlN chip. (b) Measured transmission of resonator 1 at TE polarization. Right-bottom axis indicate the resonance linewidth of corresponding TE$_{00}$ (line + circles) and TE$_{10}$ (line + triangles) modes. Different longitudinal modes with relative mode number $\mu$ are marked in the green rectangle. (c) An image of the MRRS array after dry etching. (d) Resonance profile of TE$_{00}$ mode near 1565.7 nm with Lorentz fitting. Inset: a microscope image of an MRR.

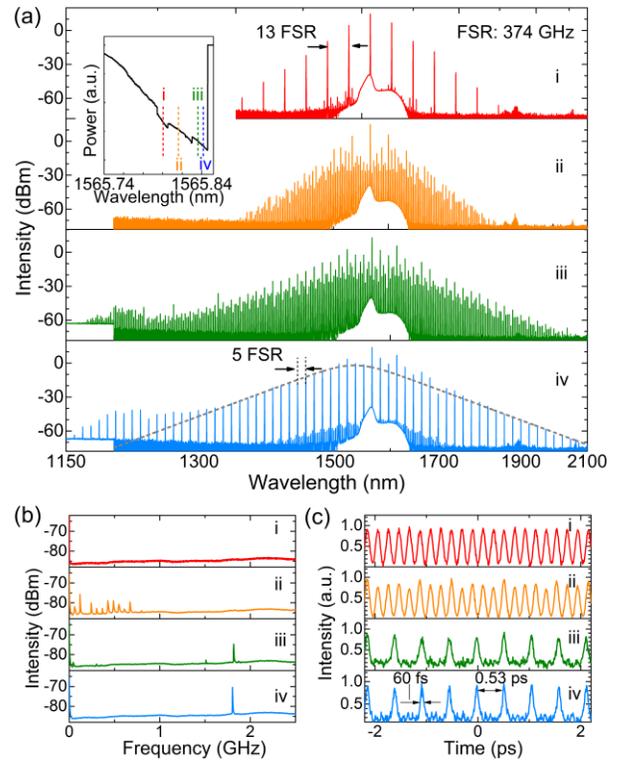

**Fig. 2.** PSC generation in resonator 1. Inset of (a): measured transmission when scanning the laser over the resonance at 1 nm/s speed and 250 mW on-chip power. Four different detuning positions (i)-(iv) are marked with dash lines. Corresponding (a) four evolution frequency comb spectra, (b) low-frequency intensity-noise measured by an electrical spectrum analyzer and (c) pulse traces measured by the autocorrelator.

The experimental setup is similar to ref. [32] and can be seen in Supplement 1. To produce PSCs, we scanned the laser wavelength at a speed of 1 nm/s speed and an on-chip power of 250 mW. A triangular resonance shape was recorded [see the inset of Fig. 2(a)] with several step-structures indicating different comb states. At the marked detuning positions (i)-(iv), we observe the formation of a primary comb, modulation instability (MI) comb, disordered SC, and 5-PSC states in sequence. The relevant optical spectra, corresponding intensity noise and measured autocorrelation pulse traces are shown in Figs. 2(a)-2(c). The disordered and perfect SCs, occurred in the final step of the transmission, have a much wider spectral range (1150-2100 nm) than those of the primary and MI comb due to the mode-locking with the presence of dispersive wave (DW) emission at ~1200 nm. The disordered SC state is a crystallization procedure related to the passage through a soliton breathing stage [24]. The PSC has a comb line spacing of 5×FSR (~1.87 THz), while all the other comb lines are suppressed significantly, suggesting a high regularity of the soliton pulse arrangement and the absence of defects. The 3-dB bandwidth of the spectrum fitted with the sech² envelope is estimated as 9 THz, which corresponds to a ~36-fs pulse. This transition from chaotic to locked states can be verified by the drastic reduction of low-frequency intensity noise. To further confirm the mode-locking, we characterized the temporal characteristics through a second-harmonic generation-based autocorrelation measurement. As show in Fig. 2(c), the pulse traces of the primary comb and MI comb are sinusoidal and have a period of 0.2 ps, inversely proportional to the comb line spacing of 13×FSR. The pulse traces of the disordered SC and PSC states have a period of 0.53 ps ($\propto$ 5×FSR), while the latter one has a higher signal to noise ratio and a narrower pulse width of 60 fs. The difference between the estimated and measured pulse width is caused by the low responsivity beyond 1600 nm of our autocorrelator. To achieve a typical single DKS with a repetition rate of 1.87 THz, one would need to exploit a much smaller microring with radius of 12 μm, which is still challenging due to the large bending losses and limitations on dispersion management in such MRRs. The primary comb results from a 23-μm-radius AlN MRR are described in Supplement 1.

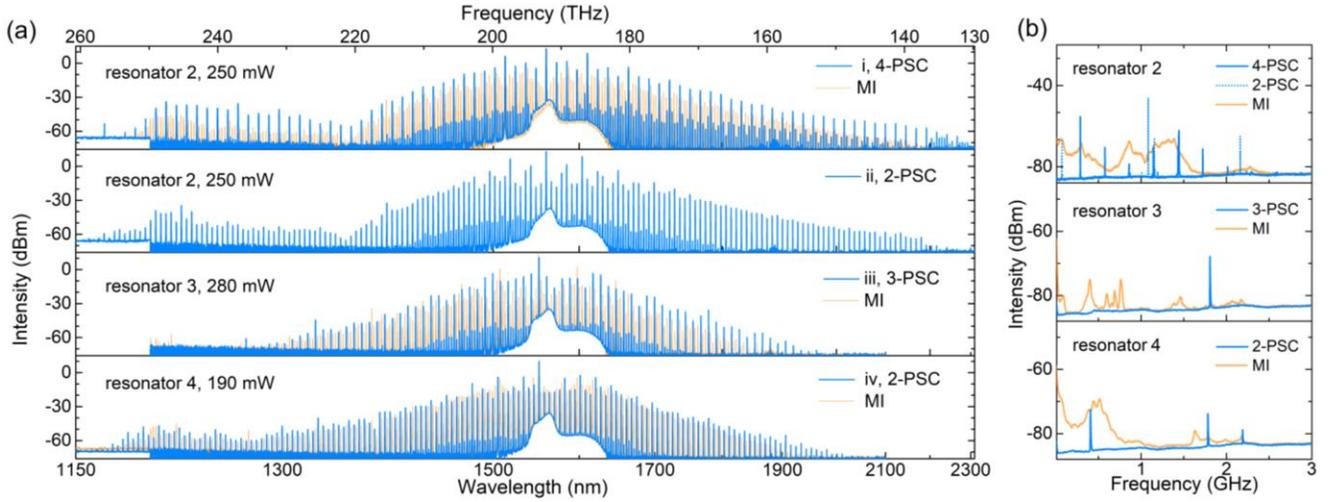

**Fig. 3.** (a) Optical spectra of PSC states with (i) $m=4$, (ii) $m=2$ for the same resonance in resonator 2, (iii) $m=3$ in resonator 3, and (iv) $m=2$ in resonator 4. The pump wavelength and on-chip powers are 1559.9 nm (250 mW), 1551.6 nm (280 mW), and 1551.7 nm (190 mW), respectively. The orange spectra are the corresponding MI combs. (b) Low-frequency intensity-noise of the microcombs in (a).

PSC spectra with different numbers are shown in Fig. 3(a), obtained from the other MRRS with the same geometry. The MI comb spectra spaced by a single-FSR are also plotted for comparison. For a resonance in the pulley-waveguide coupled resonator 2, by forward tuning the pump wavelength, we observe the near octave-spanning (1170-2300 nm) 4-PSC (i) and 2-PSC (ii) successively. The dips near 1350 nm in the comb profiles results from the antiphase-matched frequencies [34] because of the pulley coupling, which was demonstrated to improve the extraction efficiency of the microcombs at short wavelength compared to the straight coupling. Moreover, a wider coupling gap $G$ of 600 nm was utilized to increase the coupling efficiency at longer wavelength [33]. Therefore, the PSCs spectra obtained from resonator 2 are wider than the others presented in this work. Considering the various external coupling parameters and fabrication variations, the real dispersion profiles and mode coupling situations are different for the MRRs with same design. The 3-PSC and 2-PSC were also obtained from resonator 3 and 4, respectively, with a total span of ~83 and ~105 THz. As shown in Fig. 3(b), all the MI microcombs have broadband beatnote noise while the PSCs feature a few sharp lines, illustrating the low noise state.

Recently, it was shown that the versatile PSC states generated in $LiNbO_3$ can be switched by tuning the pump frequency in a bi-directional way due to the photorefractive effect in this material [35]. Here, a transition from 4-PSC to 2-PSC was also obtained in AlN by simply red-tuning the pump wavelength. By comparing the resonance characteristics with the PSC results (see Supplement 1), we can find the PSC number is tightly linked to the AMX position with respect to the pump resonance. In addition to the AMXs, the modulated wave can also be induced by other effects such as Kelly sidebands, birefringence [36], and nonlinear mode coupling [25], which have been theoretically and experimentally investigated.

In addition to decreasing the propagation loss, another degree of freedom to extend the soliton crystal bandwidth to an octave-spanning is engineering the geometry dispersion to obtain a desirable DW wavelength. A series of simulated $D_{int}$ curves [28] versus top width $RW$ of MRRs are presented in Fig. 4(a) with a fixed radius of 60 μm, sidewall angle of 72°, and thickness of 1.2 μm. $D_{int}$ is equal to 0 at 1155 nm when the $RW$ = 2.29 μm, while the position blue-shifts to 1105 and 1060 nm when $RW$ is 2.2 and 2.1 μm. We simulated the 2-PSC spectra [Fig. 4(b)] using a Lugiato–Lefever equation solver [37] with the calculated dispersion and $Q$ parameters extracted from the target resonance in resonator 4. At $RW$ = 2.29 μm, the simulated PSC spectrum agrees the experimental result very well with a tolerable deviation of the DW wavelength. The inset shows the simulated pulse (~28 fs width), which is uniformly distributed with a period of half of the cavity round-trip time. Owing to the blue-shift of the DW at $RW$ = 2.2 μm, the PSC spectrum (1056-2165 nm) exceeds an octave-span. For the 2.1-μm-wide MRR, even though the DW wavelength is shorter, the higher normal dispersion at low wavelength limits the accessible comb lines. We believe that octave-spanning PSCs are attainable using an appropriate ring width and an optimized external coupling scheme. Moreover, an octave-spanning single-DKS was demonstrated in AlN MRRs with dual DW emission, which can be achieved by optimizing the ring thickness [29].

In summary, we have demonstrated broadband perfect soliton crystals ($m$ = 2, 3, 4, 5) with THz repetition rates (~0.75, ~1.12, ~1.5, ~1.87 THz) from AlN MRRs with same geometry but different coupling. The low anomalous dispersion and the optimized external coupling enable efficient extraction of the near octave-spanning (1170-2300 nm) PSCs whose soliton number can be switched from 4 to 2. The coherence of the soliton crystals can be confirmed by the low intensity noise and the compressed pulse (60-fs width for 5-PSC). Meanwhile, further investigation is needed to realize more arbitrarily tunable PSC states in a single resonance with fewer pump variables. Versatile switching of a soliton crystal has been demonstrated in a $Si_3N_4$ MRR by adjusting the voltage of an integrated graphene layer [38]. Deterministic generation and switching of DKS was also realized in a thermally controlled MRR

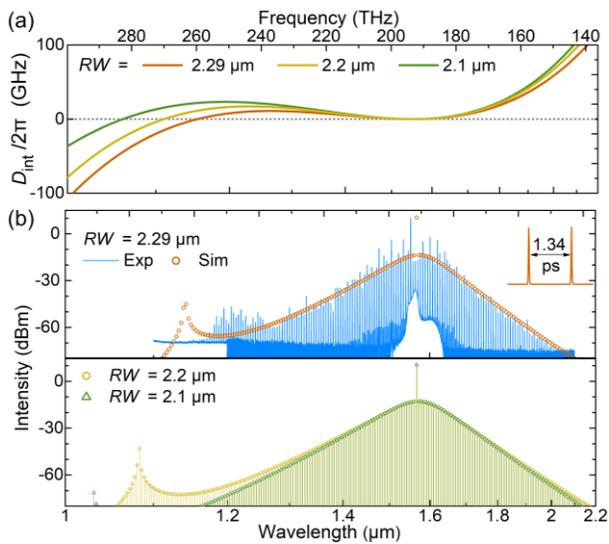

**Fig. 4.** (a) Simulated $D_{int}$ curves for 60-μm-radius microrings with fixed thickness of 1.2 μm, sidewall angle of 72°, and different widths of 2.29 (red), 2.2 (yellow), and 2.1 μm (green). (b) Simulated 2-PSC soliton output spectral (230 mW) in comparison with the measured spectrum [blue].

[39]. These are helpful for realizing switchable octave-spanning PSCs in a single AlN MRR in the future.

The demonstration of such near-octave PSCs with various repetition rates should significantly enhance and extend the application potential of soliton microcombs. For example, we can expect that the generation of coherent THz wave and arbitrary waveform, as well as with dual-PSC generated from mixed polarization. Additionally, there is also the possibility to realize the self-referencing soliton microcombs in a single AlN chip by combining the Pockels and Kerr nonlinearities [29].

**Funding.** Science Foundation Ireland (17/NSFC/4918); National Natural Science Foundation of China (61861136001).

**Disclosures.** The authors declare no conflicts of interest.

See Supplement 1 for supporting content.

## REFERENCES

1. T. Herr, V. Brasch, J. D. Jost, C. Y. Wang, N. M. Kondratiev, M. L. Gorodetsky, and T. J. Kippenberg, Nat. Photonics **8**, 145 (2014).
2. X. Yi, Q.-F. Yang, K. Y. Yang, M.-G. Suh, and K. Vahala, Optica **2**, 1078 (2015).
3. V. Brasch, M. Geiselmann, T. Herr, G. Lihachev, M. H. P. Pfeiffer, M. L. Gorodetsky, and T. J. Kippenberg, Science **351**, 357 (2016).
4. C. Joshi, J. K. Jang, K. Luke, X. Ji, S. A. Miller, A. Klenner, Y. Okawachi, M. Lipson, and A. L. Gaeta, Opt. Lett. **41**, 2565 (2016).
5. Q. Li, T. C. Briles, D. A. Westly, T. E. Drake, J. R. Stone, B. R. Ilic, S. A. Diddams, S. B. Papp, and K. Srinivasan, Optica **4**, 193 (2017).
6. M. H. P. Pfeiffer, C. Herkommer, J. Liu, H. Guo, M. Karpov, E. Lucas, M. Zervas, and T. J. Kippenberg, Optica **4**, 684 (2017).
7. X. Ji, J. K. Jang, U. D. Dave, M. Corato-Zanarella, C. Joshi, A. L. Gaeta, and M. Lipson, Laser & Photonics Rev. **15**, 2000353 (2021).
8. Z. Gong, A. Bruch, M. Shen, X. Guo, H. Jung, L. Fan, X. Liu, L. Zhang, J. Wang, J. Li, J. Yan, and H. X. Tang, Opt. Lett. **43**, 4366 (2018).
9. Y. He, Q.-F. Yang, J. Ling, R. Luo, H. Liang, M. Li, B. Shen, H. Wang, K. Vahala, and Q. Lin, Optica **6**, 1138 (2019).
10. Z. Gong, X. Liu, Y. Xu, and H. X. Tang, Optica **7**, 1275 (2020).
11. G. Moille, L. Chang, W. Xie, A. Rao, X. Lu, M. Davanço, J. E. Bowers, and K. Srinivasan, Laser & Photonics Rev. **14**, 2000022 (2020).
12. M.-G. Suh, Q.-F. Yang, K. Y. Yang, X. Yi, and K. J. Vahala, Science **354**, 600 (2016).
13. T. J. Kippenberg, A. L. Gaeta, M. Lipson, and M. L. Gorodetsky, Science **361**, eaan8083 (2018).
14. E. Lucas, P. Brochard, R. Bouchand, S. Schilt, T. Südmeyer, and T. J. Kippenberg, Nat. Commun. **11**, 374 (2020).
15. E. Lucas, M. Karpov, H. Guo, M. Gorodetsky, and T. J. Kippenberg, Nat. Commun. **8**, 736 (2017).
16. M. Yu, J. K. Jang, Y. Okawachi, A. G. Griffith, K. Luke, S. A. Miller, X. Ji, M. Lipson, and A. L. Gaeta, Nat. Commun. **8**, 14569 (2017).
17. Q.-F. Yang, X. Yi, K. Y. Yang, and K. Vahala, Nat. Phys. **13**, 53 (2016).
18. W. Weng, R. Bouchand, E. Lucas, E. Obrzud, T. Herr, and T. J. Kippenberg, Nat. Commun. **11**, 2402 (2020).
19. P. Del'Haye, A. Coillet, W. Loh, K. Beha, S. B. Papp, and S. A. Diddams, Nat. Commun. **6**, 5668 (2015).
20. D. C. Cole, E. S. Lamb, P. Del'Haye, S. A. Diddams, and S. B. Papp, Nat. Photonics **11**, 671 (2017).
21. X. Xue, Y. Xuan, Y. Liu, P.-H. Wang, S. Chen, J. Wang, D. E. Leaird, M. Qi, and A. M. Weiner, Nat. Photonics **9**, 594 (2015).
22. E. Nazemosadat, A. Fülöp, Ó. B. Helgason, P.-H. Wang, Y. Xuan, D. E. Leaird, M. Qi, E. Silvestre, A. M. Weiner, and V. Torres-Company, Phys. Rev. A **103**, 013513 (2021).
23. W. Wang, Z. Lu, W. Zhang, S. T. Chu, B. E. Little, L. Wang, X. Xie, M. Liu, Q. Yang, L. Wang, J. Zhao, G. Wang, Q. Sun, Y. Liu, Y. Wang, and W. Zhao, Opt. Lett. **43**, 2002 (2018).
24. M. Karpov, M. H. Pfeiffer, H. Guo, W. Weng, J. Liu, and T. J. Kippenberg, Nat. Phys. **15**, 1071 (2019).
25. T. Huang, J. Pan, Z. Cheng, G. Xu, Z. Wu, T. Du, S. Zeng, and P. P. Shum, Phys. Rev. A **103**, 023502 (2021).
26. X. Liu, C. Sun, B. Xiong, L. Wang, J. Wang, Y. Han, Z. Hao, H. Li, Y. Luo, J. Yan, T. Wei, Y. Zhang, and J. Wang, ACS Photonics **5**, 1943 (2018).
27. Y. Zheng, C. Sun, B. Xiong, L. Wang, J. Wang, Y. Han, Z. Hao, H. Li, J. Yu, Y. Luo, J. Yan, T. Wei, Y. Zhang, and J. Wang, *Conference on Lasers and Electro-Optics* (Optical Society of America, 2020), paper SW4J.3.
28. H. Weng, J. Liu, A. A. Afridi, J. Li, J. Dai, X. Ma, Y. Zhang, Q. Lu, J. F. Donegan, and W. Guo, Opt. Lett. **46**, 540 (2021).
29. X. Liu, Z. Gong, A. W. Bruch, J. B. Surya, J. Lu, and H. X. Tang, arXiv:2012.13496 (2020).
30. X. Guo, C.-L. Zou, H. Jung, Z. Gong, A. Bruch, L. Jiang, and H. X. Tang, Phys. Rev. Appl. 10, 014012 (2018).
31. A. W. Bruch, X. Liu, Z. Gong, J. B. Surya, M. Li, C.-L. Zou, and H. X. Tang, Nat. Photonics **15**, 21 (2021).
32. H. Weng, J. Liu, A. A. Afridi, J. Li, J. Dai, X. Ma, Y. Zhang, Q. Lu, J. F. Donegan, and W. Guo, arxiv: 2012.10956 (2020).
33. J. Liu, H. Weng, A. A. Afridi, J. Li, J. Dai, X. Ma, H. Long, Y. Zhang, Q. Lu, J. F. Donegan, and W. Guo, Opt. Express **28**, 19270 (2020).
34. G. Moille, Q. Li, T. C. Briles, S.-P. Yu, T. Drake, X. Lu, A. Rao, D. Westly, S. B. Papp, and K. Srinivasan, Opt. Lett. **44**, 4737 (2019).
35. Y. He, J. Ling, M. Li, and Q. Lin, Laser & Photonics Rev. **14**, 1900339 (2020).
36. Y. Wang, F. Leo, J. Fatome, M. Erkintalo, S. G. Murdoch, and S. Coen, Optica **4**, 855 (2017).
37. G. Moille, Q. Li, X. Lu, and K. Srinivasan, J Res. Natl. Inst. Stan. **124**, 124012 (2019).
38. B. Yao, S. W. Huang, Y. Liu, A. K. Vinod, C. Choi, M. Hoff, Y. Li, M. Yu, Z. Feng, D. L. Kwong, Y. Huang, Y. Rao, X. Duan, and C. W. Wong, Nature **558**, 410 (2018).
39. Z. Lu, W. Wang, W. Zhang, S. T. Chu, B. E. Little, M. Liu, L. Wang, C.-L. Zou, C.-H. Dong, B. Zhao, and W. Zhao, AIP Advances **9**, 025314 (2019).

# Supplementary 1 for
# "Near octave-spanning perfect soliton crystals in AlN microresonators"

## 1. EXPERIMENTAL SETUP

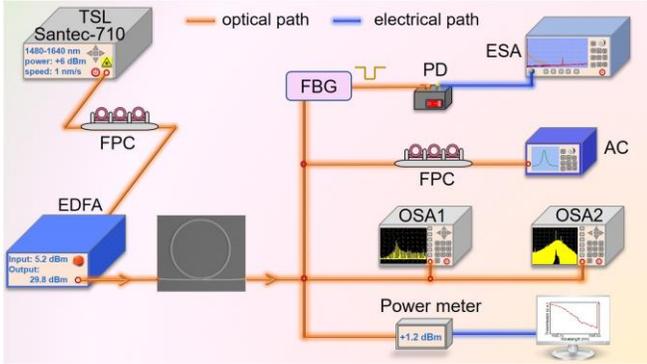

**Fig. S1.** Schematic of the experiment setup used for perfect soliton crystals generation. Component descriptions are in the text.

The experimental setup is presented Fig. S1. The light source used in the experiment is a tunable laser ranging from 1480 to 1640 nm (Santec TSL-710). A fiber polarization controller (FPC) is used to adjust the polarization. An erbium-doped fiber amplifier (EDFA) is used to amplify the laser power for pumping the resonator through a lensed fiber with a spot size of 2.5 μm. The generated microcombs are collected with another lensed fiber and divided into four branches, one of which is connected with the optical spectrum analyzers (1200-2400 nm and 600-1700 nm range) for recording the spectrum. By recording the output with a power meter (Santec MPM210) synched with the tunable laser, we can obtain the transmission with resolution of 0.1 pm. The third part is injected into a fiber Bragg grating (FBG bandwidth 0.4 nm) to suppress the pump power for intensity noise measurement via an electrical spectrum analyzer (ESA), after a photodiode (PD). Note that the results in Fig. 2(c) were measured with ESA directly because of the wavelength limitation (1545-1564 nm) of the FBG in our laboratory. The fourth branch is connected with a FPC followed by an autocorrelator.

## 2. DEVICES CHARACTERIZATION AND COMPARISON OF THE PERFECT SOLITON CRYSTALS

Figure S2(a) shows the transmission spectra of different resonators, where the $TE_{00}$, $TE_{10}$, and $TM_{00}$ modes are marked by circles, triangles, and squares, respectively. $\mu=0$ indicates the position where the different modes are closest to each other. For resonator 2 and 4, the mode coupling occurs between two fundamental modes with mixed polarizations even we pump the microresonators at specific quasi-TE or quasi-TM polarization, which has also been investigated in ref. [1]. For resonator 3, the three modes approach each other at ~1560.7 nm. Figure S2(b) plots the free spectral ranges (FSRs) and loaded $Q$ factors ($Q_{load}$) of different modes in resonator 4, where an avoided mode crossing (AMX) between $TE_{00}$ and $TM_{00}$ modes was observed at mode $\mu=0$ with the evidence of $Q_{load}$ crossing, accompanied by the obviously changed FSRs. In addition, the FSRs of $TE_{00}$ and $TE_{10}$ modes tend to approach each other at 1570 nm, illustrating another mode coupling. Therefore, diverse mode interactions in the telecommunication band can be realized repeatably in these multiple AlN microresonators with the same geometry. In the experiment, for resonators 2, 3 and 4, we pumped the $TE_{00}$ modes that belong to mode $\mu=-4$, $\mu=3$, and $\mu=2$ to generate the perfect soliton crystal (PSC) with number of 4, 3, and 2. It can be seen that the PSC number is same as the AMX position with respect to the pump resonance. However, the versatile switching among various PSC states [2] and a previous report [3] imply that there are still other factors affecting the PSC number such as multiple modal crossings and nonlinear mode coupling [4]. Table S1 summarizes the design parameters and resonance characteristics of different devices, as well as the pump conditions and PSC results. $Q_{int}$ is the intrinsic $Q$ factor.

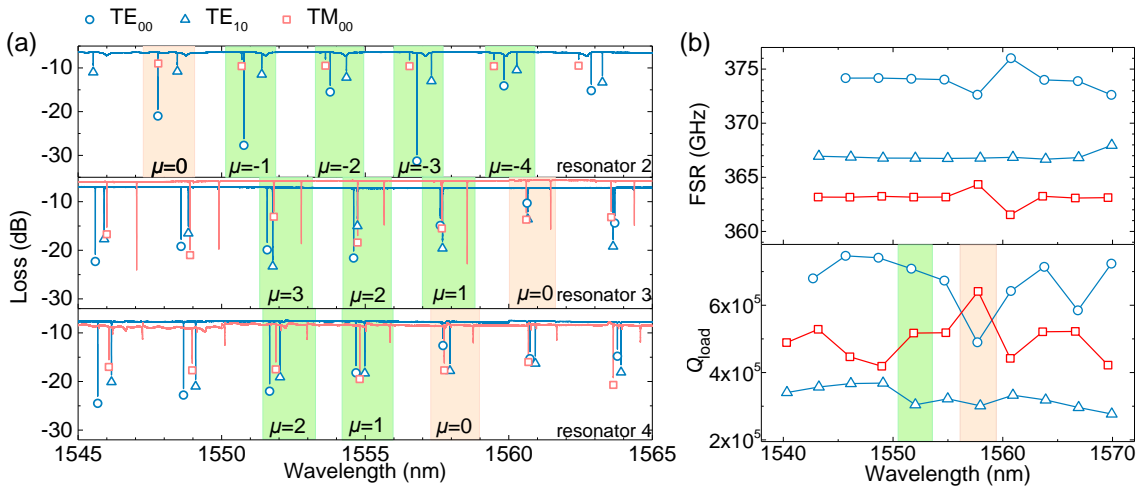

**Fig. S2.** (a) Transmission spectra of resonator 2, 3, and 4 (from top to bottom). The $TE_{00}$, $TE_{10}$, and $TM_{00}$ modes are marked by circles, triangles, and squares, respectively. Relative mode number $\mu=0$ indicate the mode coupling position, while the $TE_{00}$ modes in the edge green rectangle are pumped to produce perfect soliton crystals. (b) Extracted loaded $Q$ factors and FSRs of the different modes in resonator 4.

Table S1: Summary of the geometry design, resonance characteristics of the resonators, pump parameters and the soliton crystal result.

| Resonator | Waveguide width (μm) | Coupling gap (μm) | Pump resonance (nm) | $Q_{load}$ | $Q_{int}$ | On-chip power (mW) | PSC number | PSC range (nm) |
|---|---|---|---|---|---|---|---|---|
| 1 | 0.9 | 0.5 | ~1565.7 | $5.2 \times 10^5$ | $1.3 \times 10^6$ | 250 | 5 | 1150-2100 |
| 2 | 1.1 | 0.6 | ~1559.9 | $4.3 \times 10^5$ | $1.7 \times 10^6$ | 250 | 4, and 2 | 1170-2300 |
| 3 | 0.9 | 0.5 | ~1551.6 | $7 \times 10^5$ | $1.9 \times 10^6$ | 280 | 3 | 1280-1930 |
| 4 | 1.1 | 0.5 | ~1551.7 | $7 \times 10^5$ | $1.8 \times 10^6$ | 190 | 2 | 1180-2000 |

## 3. PRIMARY COMBS GENERATED FROM A SMALLER MICRORESONATOR

In this section, we present the primary comb results generated from a smaller microring with a radius of 23 μm (FSR ~1 THz), and a cross section of 2.3×1.2μm². Figure S3(a) shows the normalized transmission of the target mode with a $Q_{load}$ and $Q_{int}$ of ~2.7×10⁵ and ~3.3×10⁵, corresponding to a propagation loss of ~1.1 dB/cm. The large bending loss will increase the threshold power and limit the accessible frequency comb bandwidth. The parameter oscillation threshold power of the resonance is measured to be 130 mW, under four four-wave mixing sidebands [red spectrum in Fig. S3(b)] was observed. By pumping the mode at a high on-chip power of 600 mW, a primary comb with line spacing of 4×FSR and spectral range of 1380-1900 nm was obtained. Therefore, it is still challenging to achieve a single-soliton microcomb with THz repetition rate in AlN.

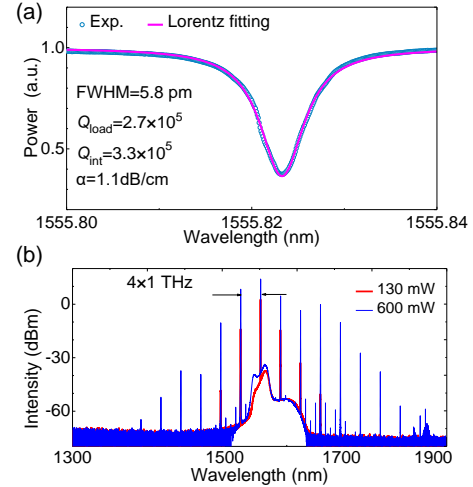

**Fig. S3.** (a) Normalized transmission of one resonance in a 23-μm-radius microresonator. (b) The generated primary comb spectra when pumping at 130 and 600 mW.

## REFERENCES

1. S. Ramelow, A. Farsi, S. Clemmen, J. S. Levy, A. R. Johnson, Y. Okawachi, M. R. Lamont, M. Lipson, and A. L. Gaeta, "Strong polarization mode coupling in microresonators," Opt. Lett. **39**, 5134-5137 (2014).
2. Y. He, J. Ling, M. Li, and Q. Lin, "Perfect Soliton Crystals on Demand," Laser & Photonics Rev. **14**, 1900339 (2020).
3. M. Karpov, M. H. Pfeiffer, H. Guo, W. Weng, J. Liu, and T. J. Kippenberg, "Dynamics of soliton crystals in optical microresonators," Nat. Phys. **15**, 1071-1077 (2019).
4. T. Huang, J. Pan, Z. Cheng, G. Xu, Z. Wu, T. Du, S. Zeng, and P. P. Shum, "Nonlinear-mode-coupling-induced soliton crystal dynamics in optical microresonators," Phys. Rev. A **103**, 023502 (2021).